\documentclass{appolb}

\usepackage{graphicx}
\usepackage{amsmath} 
\usepackage{cite}

\begin{document}

 %\eqsec  % uncomment this line to get equations numbered by (sec.num)

\title{Lattice QCD at finite temperature: some aspects \\ related to chiral symmetry\thanks{Plenary talk at Quark Matter 2022, Krak\'ow, Poland, April 4-10 2022.}
}

\author{Gert Aarts
\address{Department of Physics, Swansea University, Swansea, SA2 8PP, United Kingdom\\
European Centre for Theoretical Studies in Nuclear Physics and Related Areas (ECT*) \& Fondazione Bruno Kessler, Strada delle Tabarelle 286, 38123 Villazzano (TN), Italy
}
%\\[3mm]
%{Third Author % of different affiliation
%\address{affiliation}
%}
%\\[3mm]
%the Name(s) of other Author(s)
%\address{affiliation}
}
\maketitle

\begin{abstract}
Out of the many exciting results obtained with the lattice approach to QCD under extreme conditions, I discuss a few selected items related to chiral symmetry: the chiral condensate as an approximate order parameter, meson screening masses, and masses of baryons and mesons, including $D_{(s)}$ mesons, when approaching the crossover from the hadronic side.
\end{abstract}
  
\section{Introduction}

Lattice QCD is {\em the} first-principle method to study QCD under extreme conditions, at least along the temperature axis in the QCD phase diagram and while considering equilibrium conditions. Indeed, studies of QCD thermodynamics, including e.g.\ the equation of state and  fluctuations of conserved charges, are nowadays routinely done at the physical point while taking the continuum limit \cite{Borsanyi:2010cj,HotQCD:2014kol,Borsanyi:2011sw,HotQCD:2012fhj}.
%also in the presence of an external magnetic field. 
Extensions to nonzero baryon density have to deal with the sign problem, which can be confidently handled either at real but small quark chemical potential, or at imaginary quark chemical potential, after analytical continuation. 
%This has resulted in estimates of e.g.\ the curvature of the phase boundary.

A clear next step, going beyond thermodynamics and the phase structure, is to consider the behaviour of hadrons. Heavy quarks (charm and bottom) and quarkonium states (charmonium and bottomonium) have been in the spotlight for a long time, due to their relevance for heavy-ion phenomenology. 
%Bottom(onium) can be treated with non-relativistic QCD (NRQCD), also at finite temperature, which has some clear advantages, both computationally and for the interpretation of  numerical data. 
In this presentation I will not be able to do justice to these developments. Instead I will focus on the light quark sector, emphasing chiral symmetry and its impact beyond the identification of the transition. In particular, I will discuss three aspects:
the chiral condensate as an approximate order parameter;
mesonic screening masses;
baryon and meson masses, especially when approaching the crossover from the hadronic side.

\section{Chiral symmetry and the thermal crossover}

\begin{figure}[t]
\centerline{
$\vcenter{\hbox{\includegraphics[width=0.5\textwidth]{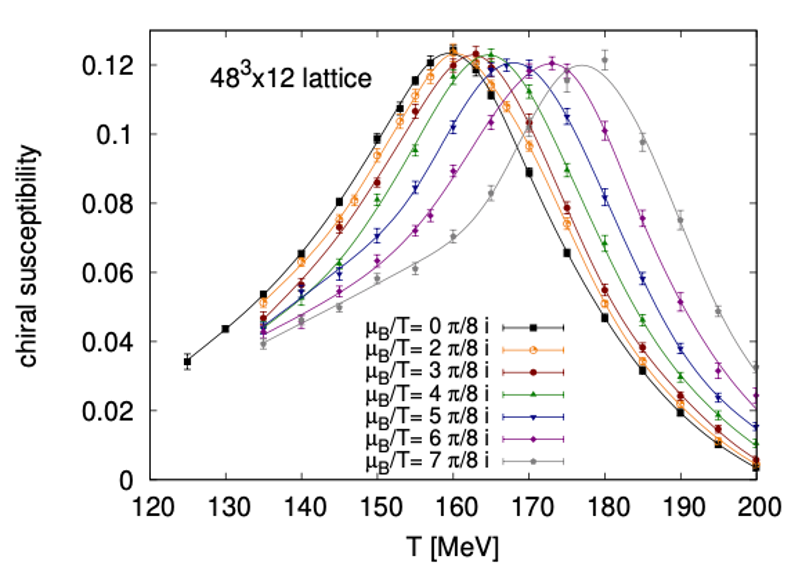}}}$
$\vcenter{\hbox{\includegraphics[width=0.5\textwidth]{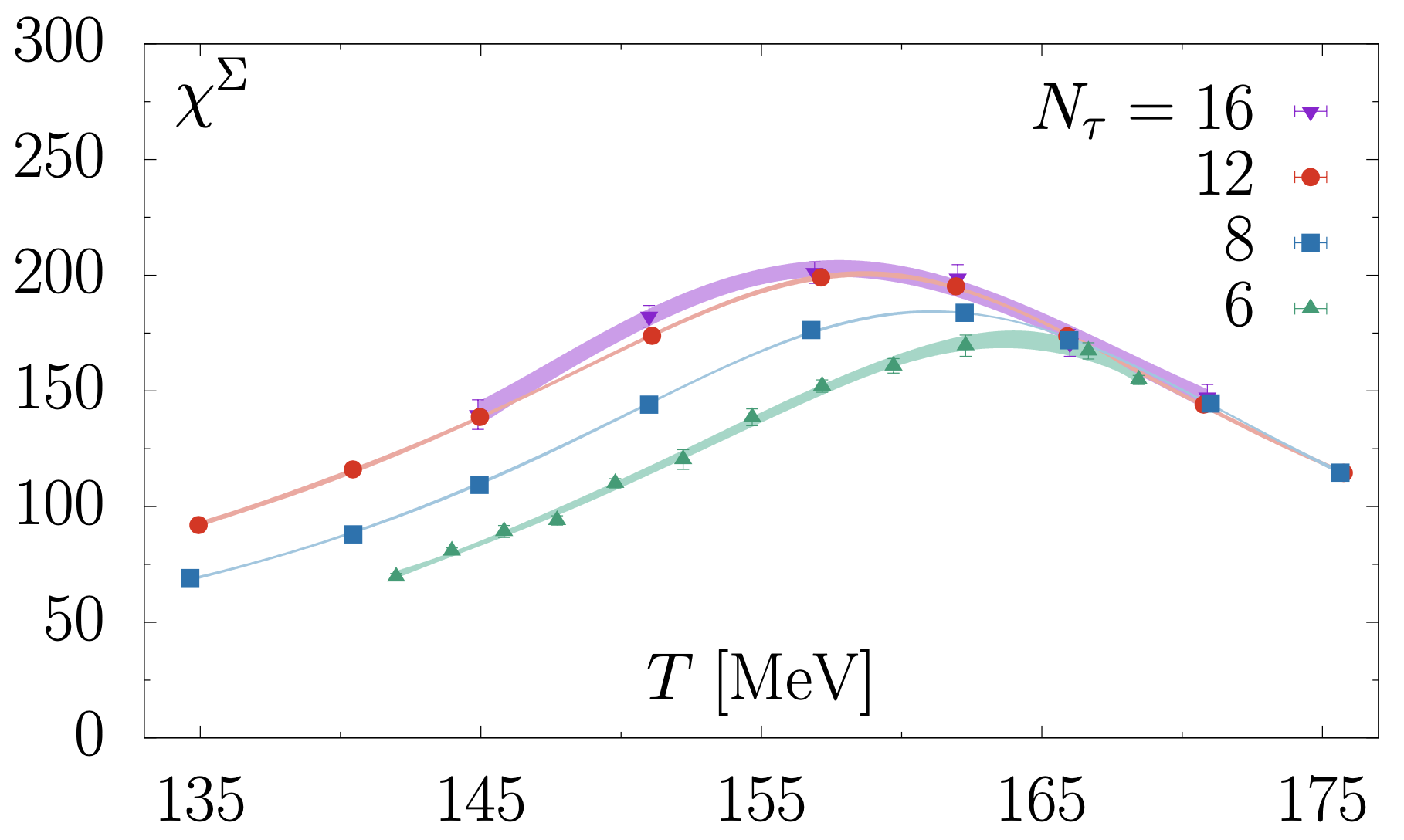}}}$
}
\caption{Chiral condensate as a function of temperature, for several values of an (imaginary) baryon chemical potential (left, Budapest-Wuppertal-Houston \cite{Borsanyi:2020fev}), and several values of the lattice spacing $a\sim 1/N_\tau$  (right, HotQCD \cite{HotQCD:2018pds}).
}
\label{fig1}
\end{figure}

\begin{figure}[t]
\centerline{
\includegraphics[width=0.6\textwidth]{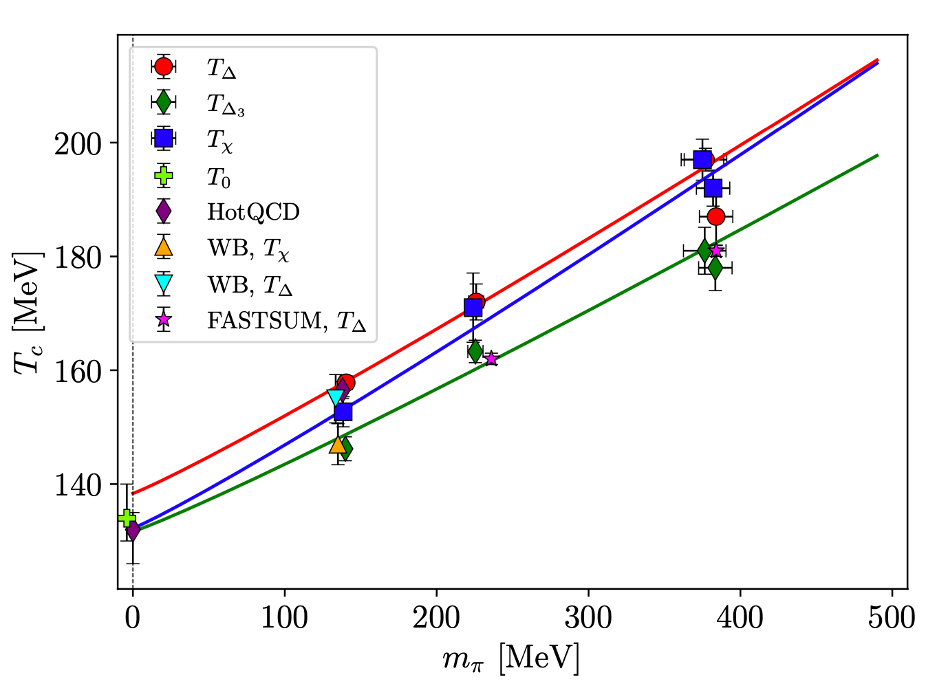}
}
\caption{Pion mass dependence of the pseudocritical temperature, obtained using twisted-mass (first three symbols), staggered (HotQCD, WB) and Wilson-clover (FASTSUM) fermions, for several quantities related to the chiral susceptibility \cite{Kotov:2021rah}.
}
\label{fig2}
\end{figure}

Chiral symmetry is an approximate symmetry, spontaneously broken in the hadronic phase.
Observables linked to chiral symmetry are used to identify the thermal crossover, notably the chiral condensate and its fluctuations, the chiral susceptibility $\chi$ (see Fig.\ \ref{fig1}). In fact, the most precise determination of the pseudocritical temperature has been obtained from the latter, with the recent values
\begin{align*}
&T_{\rm pc}^\chi = 158.0(6) \mbox{\;MeV} &&\mbox{Budapest-Wuppertal-Houston \cite{Borsanyi:2020fev}}, \\
&T_{\rm pc}^\chi = 156.5(1.5) \mbox{\;MeV} &&\mbox{HotQCD \cite{HotQCD:2018pds}}.
\end{align*}
From a lattice perspective, the results quoted here have been obtained using staggered quarks, for which simulations are carried out at the physical point and a continuum extrapolation is feasible. Wilson-type quarks are more expensive to simulate and hence results are not available at both the physical point and in the continuum limit. Yet there is steady progress here as well, as demonstrated in Fig.\ \ref{fig2}, which includes results from twisted-mass  \cite{Kotov:2021rah} and Wilson-clover \cite{Aarts:2020vyb} fermions. These are not both at the physical point (i.e.\ $m_\pi> m_\pi^{\rm nature}$) and in the continuum limit; nevertheless, an extrapolation in $m_\pi$ using Wilson-type quarks only yields $T^{\bar\psi\psi}_{\rm pc}=158(3)$ MeV at the physical point  \cite{Aarts:2020vyb}, in agreement with the numbers quoted above.

\section{Mesonic screening masses}

Going beyond thermodynamics, one may ask how chiral symmetry restoration affects hadrons and quarks. As a first step, we consider meson screening masses (or spatial screening lengths). These are computable on the same lattices as used for thermodynamics, since the computation benefits from large spatial volumes, while the Euclidean time is integrated over.
Screening masses interpolate between the pole mass at $T=0$ and $\sqrt{(2\pi T)^2+m_q^2} \approx 2\pi T$ at high $T$. They can therefore be used to probe both chiral symmetry restoration as well as the perturbative high-temperature limit.

\begin{figure}[t]
\centerline{
\includegraphics[width=0.5\textwidth]{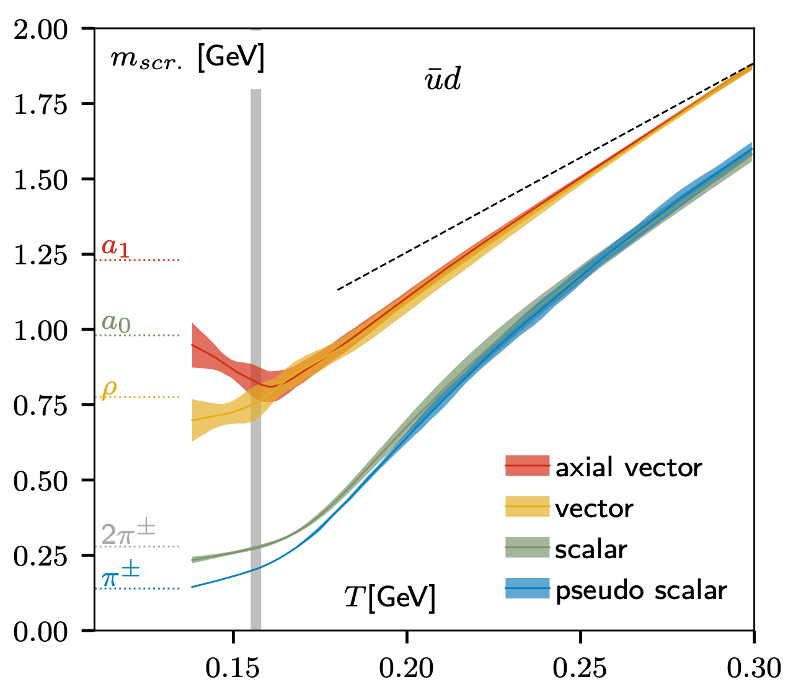} \hspace*{0.4cm}
\includegraphics[width=0.4\textwidth]{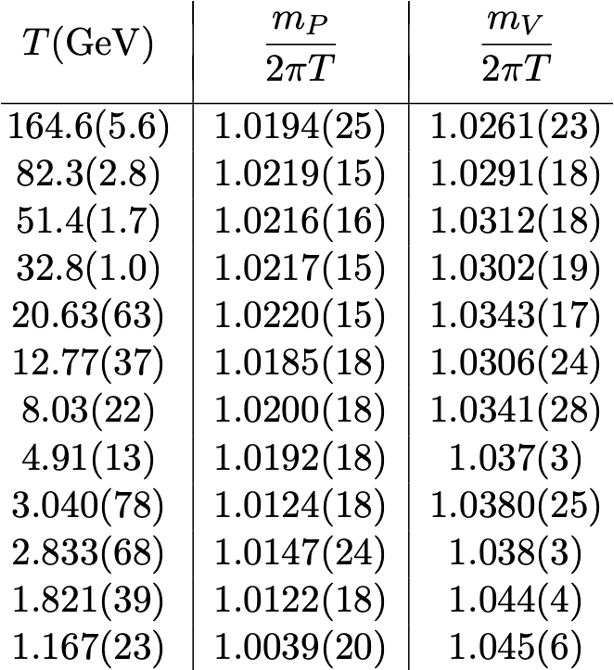}
}
\caption{
Left: screening masses as a function of temperature, for $\bar u d$ mesons \cite{Bazavov:2019www}. Right: screening masses normalised with $2\pi T$, up to {\em very} high temperature \cite{DallaBrida:2021ddx}.
}
\label{fig3}
\end{figure}

This is in demonstrated in Fig.\ \ref{fig3}: screening masses in the light-quark sector are shown on the left \cite{Bazavov:2019www}, with in particular the degeneracy in the vector ($\rho$) and axial-vector ($a_1$) channels emerging at or close to the transition. Normalised screening masses up to {\em very} high temperatures (note that units are GeV) are given on the right, including small but numerically significant deviations from $2\pi T$, which can be analysed in perturbation theory \cite{DallaBrida:2021ddx}.

\section{Baryon and meson masses}
\label{sec4}

Phenomenologically, the interest lies not so much in screening masses but rather in hadron masses. Ideally, information on the spectrum and in-medium modification is extracted from spectral functions $\rho(\omega, \mathbf{p}; T)$. Unfortunately, the inversion of the integral relating the Euclidean correlator and the spectral function is a classic ill-posed problem. 
The next best thing is to start with the assumption that hadrons exist at low temperature, with a computable mass for the groundstates, and subsequently raise the temperature of the hadron gas to follow the response. Groundstate masses are extracted from temporal lattice correlators, which benefits from having access to many Euclidean time points. This motivates the use of anisotropic lattices, with $a_\tau\ll a_s$, leading to quite different lattice geometries compared to the thermodynamic studies. 

In the following I will report on work by the FASTSUM collaboration \cite{Aarts:2020vyb}. Briefly, FASTSUM uses $N_f=2+1$ flavours of Wilson-clover quarks, 
%for ease of spectroscopy, 
with an anisotropy of $a_s/a_\tau\sim 3.45$ and a lattice cutoff of $a_\tau^{-1}\sim 6$ GeV. Light quarks are heavier than in nature;  results are shown for two sets of ensembles: Generation 2 with $m_\pi=384(4)$ MeV and Generation 2L with $m_\pi=236(2)$ MeV. In Gen2 (2L) we have 4(5) ensembles in the hadronic phase and
5(6) in the quark-gluon plasma. The transition temperature as determined by the chiral susceptibility is 170(4) and 165(3) MeV respectively.  

\begin{figure}[t]
\centerline{
\includegraphics[width=0.5\textwidth]{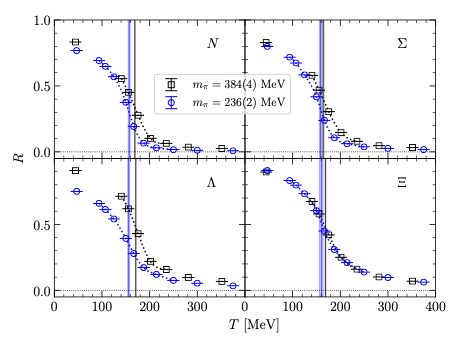} 
\includegraphics[width=0.5\textwidth]{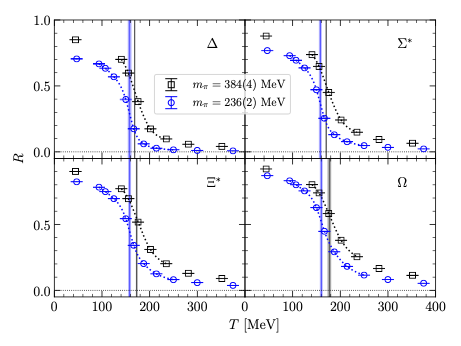}
}
\caption{
Parity-doubling parameter $R$ as a function of temperature for octet (left) and decuplet (right) baryons, for two sets of ensembles labelled by the pion mass. The vertical lines indicate the inflection points, which shift to lower temperature as the light quarks get lighter
 \cite{Aarts:2020vyb}.
  }
\label{fig4}
\end{figure}

Let us start with baryons \cite{Aarts:2015mma,Aarts:2017rrl,Aarts:2018glk}. Chiral symmetry breaking is manifested here by the absence of parity doubling, i.e.\ in vacuum 
positive- and negative-parity states are nondegenerate.\footnote{Consider e.g.\ the nucleons $m_N=m_+ = 939$ MeV and $m_{N^*}=m_- = 1535$ MeV.}  One can show that if chiral symmetry is unbroken, parity doubling is present, not only for states, but already at the level of Euclidean correlators  \cite{Aarts:2017rrl}. Hence this can be investigated without any further assumptions on the spectral content. Denoting the positive/negative-parity correlator with $G_{+/-}$, we may construct a quasi-order parameter
\[
R=\sum_\tau \frac{G_+(\tau) - G_-(\tau)}{G_+(\tau) + G_-(\tau)},
\]
which interpolates between 1 at low temperature---assuming that  the groundstates $m_\pm$ dominate and that $m_-\gg m_+$---and 0 when chiral symmetry is restored.
This $R$ parameter is shown in Fig.\ \ref{fig4} for octet and decuplet baryons, for both sets of ensembles  \cite{Aarts:2020vyb}. We observe that indeed $R$ changes from about one to close to zero, with the inflection points shifting to lower temperature as the light quarks get lighter. The inflection-point temperatures for the ensembles with the lighter pion lie between 157 and 160 MeV, close to the thermal crossover.

\begin{figure}[t]
\centerline{
\includegraphics[width=0.5\textwidth]{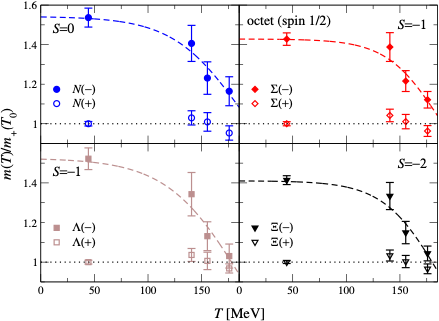} 
\includegraphics[width=0.5\textwidth]{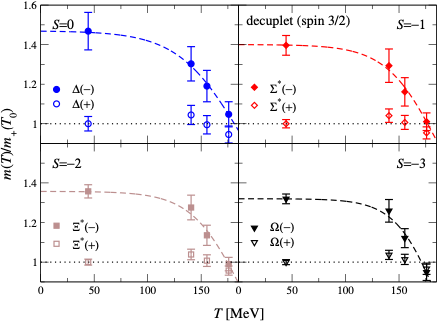}
}
\caption{
Temperature dependence of the groundstate masses, normalised with $m_+$ at the lowest temperature, in the hadronic phase, for octet (left) and decuplet (right) baryons, for the Gen2 ensembles.
Positive- (negative-) parity masses are indicated with open (closed) symbols
 \cite{Aarts:2018glk}.
 }
\label{fig5}
\end{figure}

The analysis can be taken a step further, by extracting the groundstate masses from the lattice correlators, for both parities. The results for the ensembles with the heavier pion are shown in Fig.\ \ref{fig5}, in the hadronic phase. Note that they are normalised with $m_+$ at the lowest temperature. We observe that the positive-parity masses are mostly independent of temperature, while the negative-parity masses are reduced, such that an approximate degeneracy emerges at the thermal crossover (recall that $T_{\rm pc}=170$ MeV for these data). While the degeneracy is expected, the manner in which this is realised can only be found by an actual computation. 
These results can be used as a benchmark for effective parity-doublet models used at finite density, see e.g.\ Ref.\ \cite{Marczenko:2018jui}.

\begin{figure}[t]
\centerline{
\includegraphics[width=0.5\textwidth]{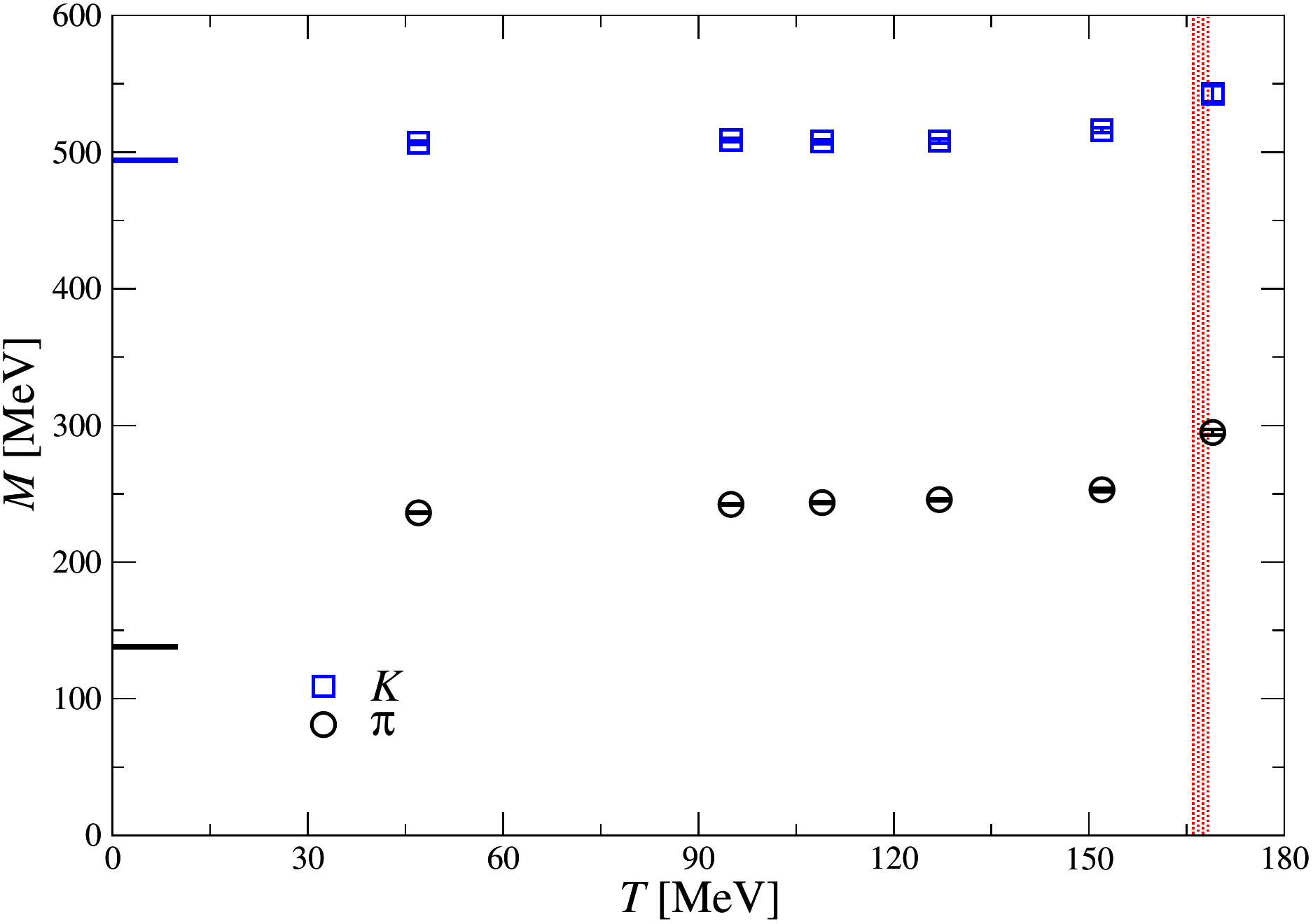} 
\includegraphics[width=0.5\textwidth]{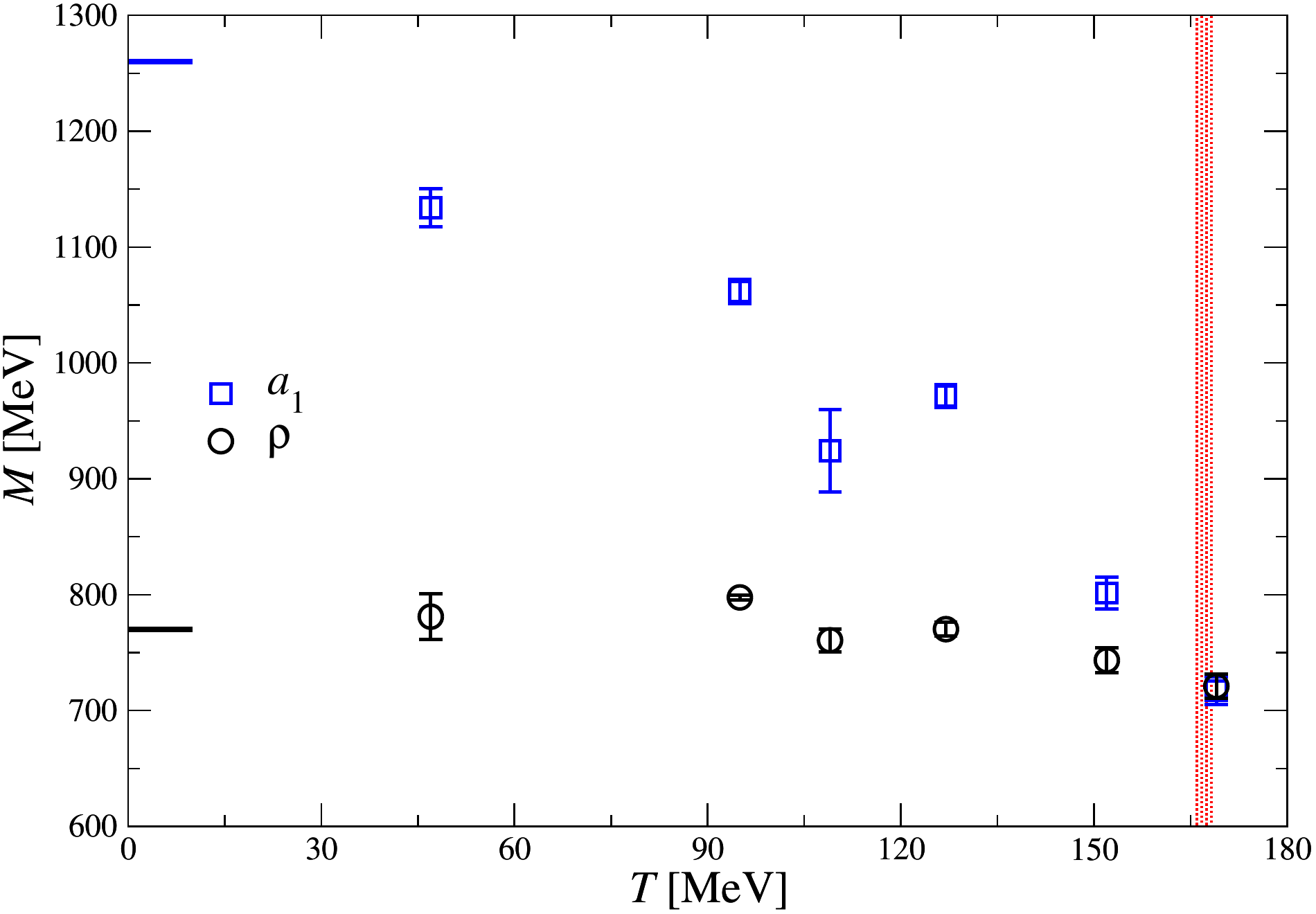}
}
\caption{Temperature dependence of the pion and kaon  (left) and the $\rho$ and $a_1$ (right) groundstate in the hadronic phase, for the Gen2L ensembles. The vertical band indicates the thermal transition, while the horizontal stubs at $T=0$ represent the PDG value; note that the light quarks are heavier than in nature, which mostly affects the pion \cite{Aarts:2022krz,inprep}.
}
\label{fig6}
\end{figure}

\begin{figure}[t]
\centerline{
\includegraphics[width=0.5\textwidth]{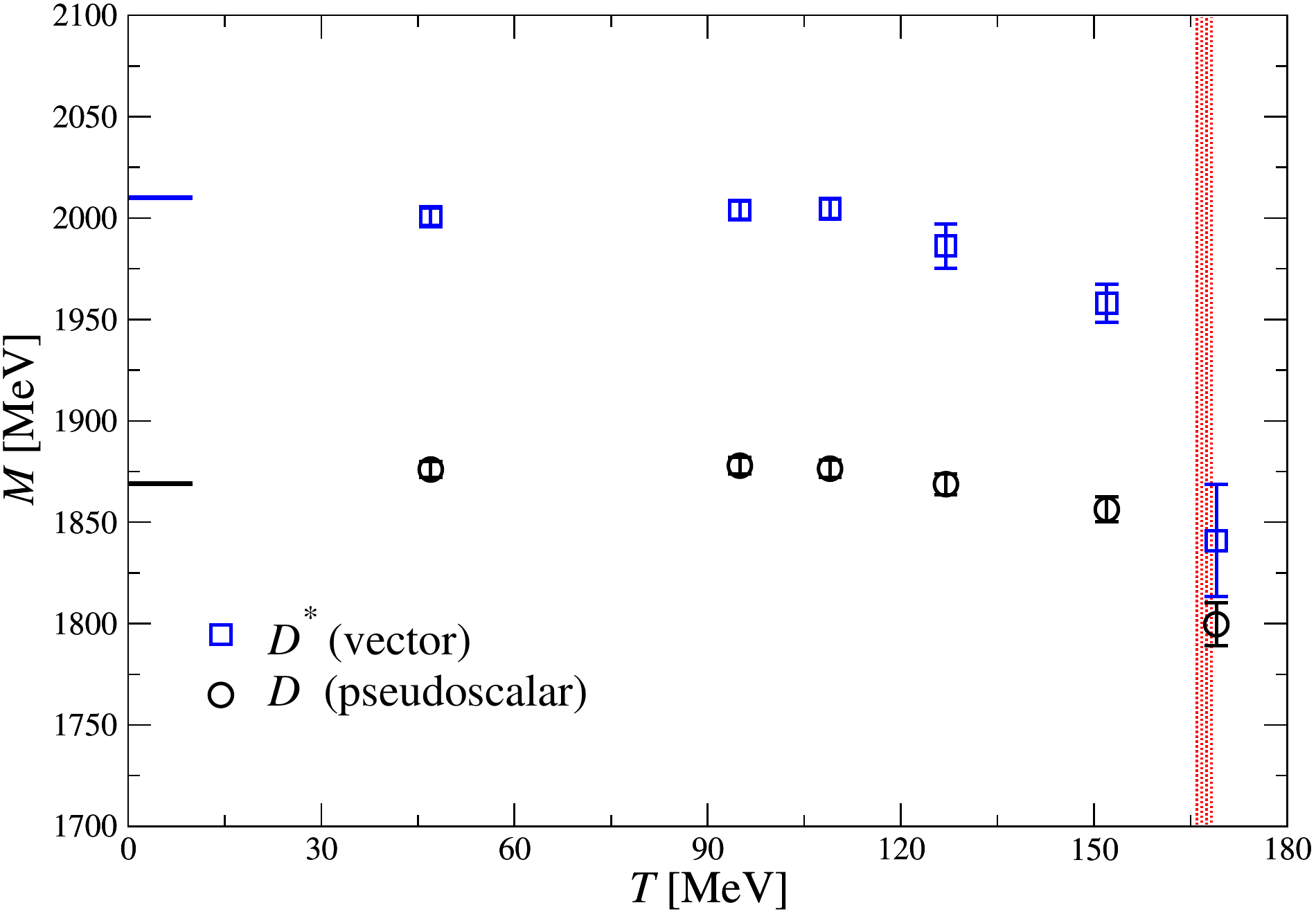} 
\includegraphics[width=0.5\textwidth]{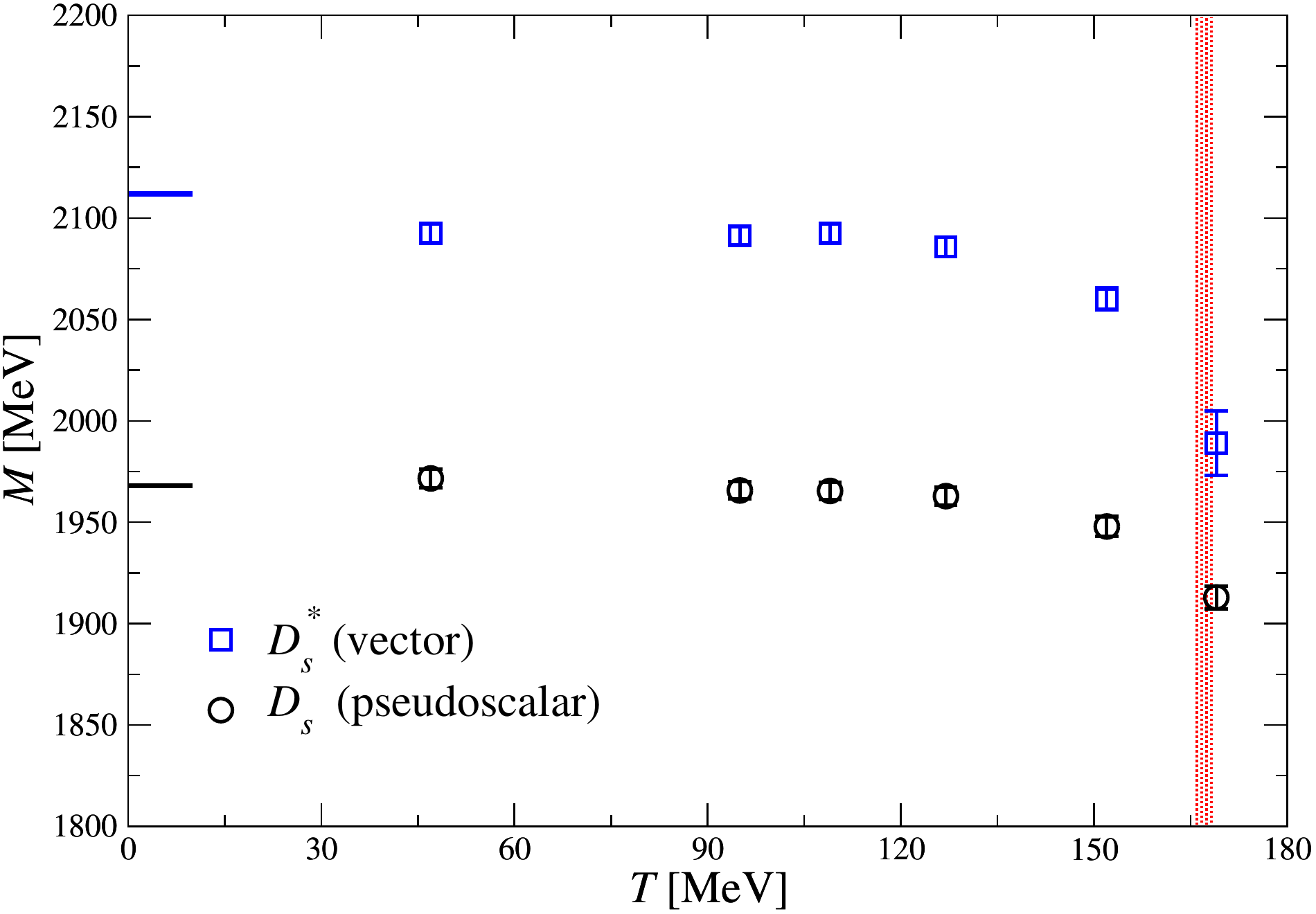}
}
\caption{As in Fig.\ \ref{fig6} for the $D$ and $D^*$ (left) and the $D_s$ and $D_s^*$ (right) mesons \cite{Aarts:2022krz}.
 }
\label{fig7}
\end{figure}

Moving now to the light meson sector, I present here some preliminary FASTSUM results obtained by assuming that spectral broadening can be ignored at low temperature. This assumption can be investigated by a detailed comparison of correlators at different temperatures \cite{Aarts:2022krz,inprep}. Results for the temperature dependence of the pion and kaon groundstates are shown in Fig.\ \ref{fig6} (left) \cite{Aarts:2022krz}. The light quarks are heavier than in nature, which mostly affects the pion.
Note that the masses of both states appear to increase slightly as the thermal crossover is approached. 
On the right the temperature dependence of the $\rho$ (vector) and $a_1$ (axial-vector) groundstates is shown \cite{inprep}. The $\rho$ state is approximately independent of temperature, while the $a_1$ state shows a rather strong temperature dependence. 
A degeneracy is a signal of SU(2)$_{\rm L} \times$ SU(2)$_{\rm R}$ chiral symmetry restoration, which is seen to occur at or very close to the thermal transition, not dissimilar as in the case of screening masses, see Fig.\ \ref{fig3} (left). We emphasise that these results are obtained under the assumption that narrow spectral functions can be used to describe the data, which might not be justified, especially in the axial-vector channel. It is noted that there are many model predictions for the thermal behaviour in these channels, see e.g.\ Ref.\ \cite{Jung:2016yxl} and references therein.

Finally, we present some FASTSUM results for $D$ and $D_s$ mesons, in the pseudoscalar and vector channels, see Fig.\ \ref{fig7} \cite{Aarts:2022krz}.
 In this case the masses are decreasing as the thermal transition is approached, quite similar in fact to the behaviour seen in effective models 
\cite{Montana:2020lfi}.

\section{Summary}

As is well known, chiral symmetry is of utmost importance for thermal QCD.
In lattice studies, it is used to study properties of thermal crossover, providing the most precise estimates of the  pseudocritical temperature. 
Going beyond thermodynamics, the expected restoration of chiral symmetry leads to emerging degeneracies in the spectrum, in the case of 
meson screening masses, for baryons and parity doubling, and for mesons and chiral partners. I have presented (new) lattice data in the hadronic phase, indicating a precursor to chiral symmetry restoration already in the hadronic phase. These results may be compared to effective model descriptions. This is especially relevant when the latter are extended to regions of the QCD phase diagram where lattice QCD is not directly applicable

\vspace*{0.2cm}
\noindent 
{\bf Acknowledgements} -- The work presented in Sec.\ \ref{sec4} was carried out by FASTSUM and I thank my colleagues for the fruitful collaboration.  
FASTSUM acknowledges DiRAC, PRACE and Supercomputing Wales for the use of computing resources. 
I am supported by the UKRI Science and Technology Facilities Council (STFC) Consolidated Grant No.\ ST/T000813/1.

\end{document}